\documentclass[12pt]{article}

\title{A New Vectorization Technique for Expression Templates in C++}
\author{J. Progsch, Y. Ineichen and A. Adelmann}

\usepackage{graphicx}
\usepackage{float}
\usepackage{cite}
\usepackage{url}
\usepackage{tikz}
\usetikzlibrary{shapes}
\usetikzlibrary{arrows,automata}

\floatstyle{ruled}
\newfloat{program}{thp}{lop}
\floatname{program}{Listing}

\begin{document}

\maketitle

\begin{abstract}
Vector operations play an important role in high performance computing and are
typically provided by highly optimized libraries that implement the BLAS (Basic Linear Algebra Subprograms) 
interface. In C++ templates and operator overloading allow the implementation of
these vector operations as expression templates which construct custom loops
at compile time and providing a more abstract interface.
Unfortunately existing expression template libraries lack the performance
of fast BLAS(Basic Linear Algebra Subprograms) implementations. This paper presents
a new approach - Statically Accelerated Loop Templates (SALT) - to close
this performance gap by combining expression templates with an aggressive loop
unrolling technique. Benchmarks were conducted using the Intel C++ compiler and
GNU Compiler Collection to assess the performance of our library
relative to Intel's Math Kernel Library as well as the Eigen template library.
The results show that the approach is able to provide optimization comparable
to the fastest available BLAS implementations, while retaining the convenience
and flexibility of a template library.
\end{abstract}

\section{Introduction}

Vector and matrix operations are important building blocks of numerical
computations like solving a system of linear equations or integrating differential
equations. It is therefore of great interest to provide optimized
implementations of these operations. Most of them can be trivially implemented
but these naive implementations are usually oblivious to hardware features and
limitations that affect the performance and are subject to compiler optimization
which may vary greatly across different compilers and optimization settings. The
Basic Linear Algebra Subprograms (BLAS) interface with roots in the
FORTRAN programming language has become the quasi standard for libraries that
provide such optimized linear algebra routines. Examples for such libraries are the
Automatically Tuned Linear Algebra Software (ATLAS), GOTO BLAS or Intel's Math Kernel
Libraries (MKL) ~\cite{intelmkl}. In the last decade
FORTRAN has been overtaken by C/C++ as the primary language for scientific
software and accordingly C/C++ bindings for BLAS libraries are being used to
benefit from the optimized implementations. Unfortunately this approach falls
short when language inherent features of C++ like operator
overloading or templates are being used. More modern approaches are present in expression
template libraries such as Blitz++ \cite{blitzpp}, Eigen \cite{eigenweb}
and the Portable Expression Template Engine (PETE) \cite{pete}.

One added benefit of the
expression template approach is the compile time construction of optimized
loops. These allow for example to reduce the amount of memory accesses of operations that
have to be expressed as multiple calls to BLAS routines but could be written as
a single loop, which is exactly what the expression template library does (loop
fusion). Generic expression template implementations still rely on the compiler
to apply additional optimization like the use of SIMD (Single Instruction Multiple Data)
instructions (vectorization) or loop unrolling. Explicit vectorization for expression templates
can be achieved with intrinsics which allow the emission of specific CPU
instructions without the use of inline assembler. Apart from using the right
instructions, the key to close to optimal performance is aggressive loop
unrolling and instruction ordering that minimizes pipeline stalling. Loop
unrolling can be done by template meta programming which tends produce suboptimal
register usage while the loop unrolling capabilities of the compilers are
inconsistent and still not optimal.

In section 2 and 3 we will discuss expression templates and vectorization. Section
4 is about the Statically Accelerated Loop Templates (SALT), where we explain
the ideas of our optimization technique and present implementation details.
In section 5 \& 6 the measurements are presented followed by a discussion of possibilities
for further parallelization by means of OpenMP threads and processes with MPI
(Message Passing Interface).  Our final conclusions are presented in section 7.
\section{Expression Templates}

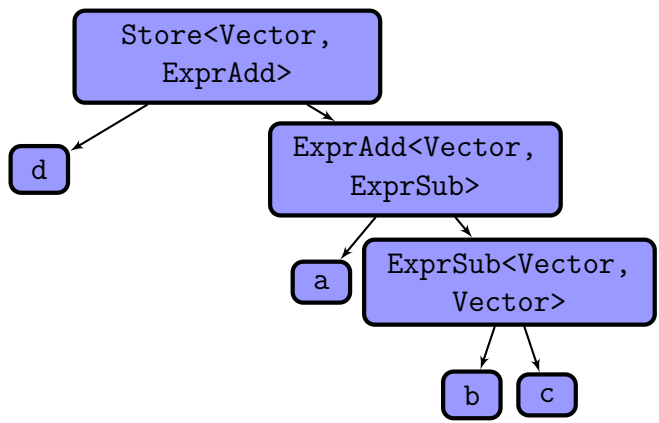
\begin{figure}
  \begin{center}

  \begin{tikzpicture}[scale=2.5,cap=round]

  \tikzstyle{statement} = [rounded corners,fill=blue!40,draw,inner
sep=1ex,text centered,ultra thick]
  \tikzstyle{decision} = [diamond,draw,fill=blue!40,text
width=4.5em,text badly centered,node distance=3cm, inner sep=0pt,ultra
thick]
  \tikzstyle{line} = [draw,-latex',thick]

  \path
      (0,0) node[text width=3.7cm,style=statement] (center)
{\texttt{Store<Vector, ExprAdd>}}

      (1,-0.6) node[text width=3.5cm,style=statement] (l1)
{\texttt{ExprAdd<Vector, ExprSub>}}

      (0.5,-1.2) node[text width=0.4cm,style=statement] (l21) {\texttt{a}}
      (1.5,-1.2) node[text width=3.5cm,style=statement] (l22)
{\texttt{ExprSub<Vector, Vector>}}

      (1.3,-1.8) node[text width=0.4cm,style=statement] (l31) {\texttt{b}}
      (1.7,-1.8) node[text width=0.4cm,style=statement] (l32) {\texttt{c}}

      (-1,-0.6) node[text width=0.4cm,style=statement] (r1)
{\texttt{d}}

    ;

    \path[line] (center) -- node [near start] {} (l1);
    \path[line] (center) -- node [near start] {} (r1);

    \path[line] (l1) -- node [near start] {} (l21);
    \path[line] (l1) -- node [near start] {} (l22);

    \path[line] (l22) -- node [near start] {} (l31);
    \path[line] (l22) -- node [near start] {} (l32);

    \end{tikzpicture}
    \caption{Expression tree for \texttt{d = a+(b-c)}}
    \label{expressiontemplate}

  \end{center}
\end{figure}

Expression templates \cite{PflaumRahimi}\cite{Hardtlein1}\cite{Hardtlein2}
were invented independently by Todd Veldhuizen
\cite{Veldhuizen95} \cite{Veldhuizen98} and David Vandevoorde.
They are an implementation technique that uses the static
evaluation abilities of modern C++ compilers together with templates to
construct a static expression tree 
at compile time as shown in figure \ref{expressiontemplate}. This is typically used for
loop fusion. This means instead of having every part of the expression individually loop
over the data and perform its operation, only a single loop is produced that
performs the composite expression. This helps reduce the loop overhead
(increments and conditional jumps), can reduce the amount memory accesses by
keeping intermediate results in registers and removes the need to allocated
temporary objects. Through the use of template meta programming it is even
possible to manipulate the expression tree at compile time to for example apply algebraic
transformations. While the nodes of the expression tree are technically
template classes their entirely static nature allows the compiler to inline
everything, in consequence the creation of instances of these classes does not incur any
overhead in the resulting machine code. This means that one can think of expression templates as a form
of code generation at compile time.

The downsides of expression templates are that they increase compile times,
the binary size and tend to produce hard to read error messages. Compile times and
binary sizes are usually of no great concern given availability of memory and the
performance of modern compilers. The error messages C++ compilers produce in
connection with intricate template constructs on the other hand are a serious
handicap. Seemingly trivial syntax errors can result in a cascade of hard
to decipher errors and warnings.

The implementation of expression templates is done by having operators and
functions return abstracted operation objects that contain all necessary information,
instead of calculating the result themselves. For example the addition
operator will return a \emph{AddExpr} object that has references to the operands
and an evaluation function. The operands in turn can also be expression objects
which means complex expressions are turned into an expression tree at compile
time. Accordingly we will refer to the expression objects as nodes.
The expression tree is only evaluated when the actual need arises (lazy evaluation).
Typically this happens in the assignment operator or in functions like the
dot product or norm. The actual loop is contained in these functions and evaluates
the expression tree for every element it needs. This also means that results
that are not requested are never computed.

\section{Vectorization}

Most modern general purpose processors offer a set of vector instructions that
operate on registers that contain multiple operands which allow instruction
level parallelism (SIMD, Single Instruction Multiple Data). These instructions
are instrumental for the optimization of vector operations since the attainable
performance is usually a multiple of what is offered by the more conservative
FPU (Floating Point Unit) instructions. To use these SIMD
instructions from C++ one has to either rely on the compiler to detect loops
that can be optimized or force their usage through non portable means
like inline assembler or intrinsics.
Since inline assembler does not have a uniform syntax across compilers it is not
well suited for use inside a template library. Intrinsics for Intel's Streaming SIMD Extension (SSE)
instruction set on the other hand integrate well into standard C++ and are
supported by most compilers.
Our template library wraps the intrinsics and the vector data types into a
\emph{vectorizer} template class, hence the actual implementations of the algorithms are
independent of the underlying instruction set. They only require a specialized
instance of the the \emph{vectorizer} template for each targeted platform.
In the present form the library only contains a specialization for Intel's SSE
instruction set.

\section{Implementation}
\subsection{General Idea}
The study of existing vectorization methods in numeric libraries and simple
experiments showed that the main performance gain of hand optimized BLAS
implementations comes from heavily unrolled loops and instruction reordering.
While it was relatively easy to have a compiler emit the ``optimal''
instructions via intrinsics we lacked the ability to unroll
the loops in a controllable fashion and even if the compiler did unroll the
loops there was no way to consistently control the relative order or multiple
sets of memory and operation instructions.

The main idea to get the required control over the loop structure was to
use abstracted loop templates. These contain calls to functions
like \emph{load},\emph{store} and \emph{operation} which are then provided
by the expression tree. This approach separates the instructions that
are executed from the order in which they are executed and gives the control
over both aspects to the library implementer, as opposed to leaving the order
to the compiler.

\begin{figure}
  \begin{center}

  \begin{tikzpicture}[scale=2.5,cap=round]

  \tikzstyle{statement} = [fill=green!40,draw,inner
sep=2ex,text centered,ultra thick]
  \tikzstyle{line} = [draw,-latex',thick, sloped, above]

\path

      (-1.2,0) node[text width=3.5cm,style=statement] (v)
{\texttt{vectorizer}}

      (1.2,0) node[text width=3.5cm,style=statement] (et)
{\texttt{expression tree}}

      (-1.2,-1.5) node[text width=3.5cm,style=statement] (lt)
{\texttt{loop template}}

      (1.2,-1.5) node[text width=3.5cm,style=statement] (ef)
{\texttt{execute function}}

    ;

    \path[line] (et) -- node {uses} (v);
    \path[line] (ef) -- node {selects} (lt);
    \path[line] (et) -- node {passed to} (ef);

    \end{tikzpicture}
    \caption{Component interaction}
    \label{components}

  \end{center}
\end{figure}
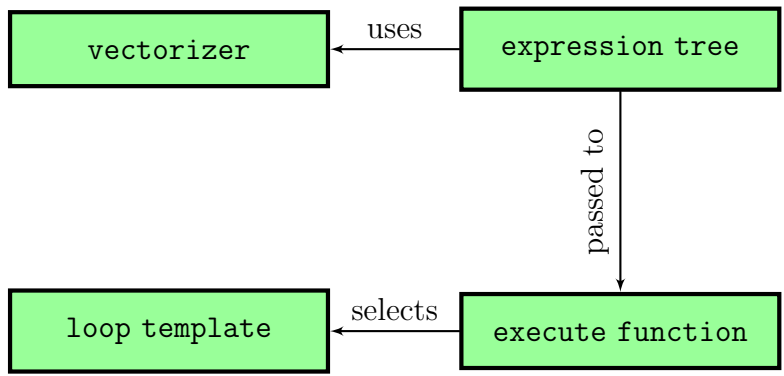

The SALT therefore has three main components which are used to assemble
the final loop (see figure \ref{components}). The \emph{vectorizer} class provides the platform specific
vector instructions, the loop templates provide the loop structure and the
expression tree defines the operation that is to be performed. These three
components are then used to assemble the actual loop inside a \emph{execute}
function that takes the expression tree as its sole argument and chooses
an appropriate loop template while the \emph{vectorizer} is implicitly chosen
by the type parameter of the participating vectors.

\subsection{Instruction Decomposition and Local Storage}
The expression tree nodes have to provide all the necessary methods and variables
that are required for the loop. The operations that are used inside the loops are
separated into \emph{load}, \emph{store} and operation. The operation functions are
further subdivided into \emph{vector\_op} and \emph{single\_op} with the second
one being used to finish up operations where the amount of data
isn't a multiple of the SIMD vector size. Additionally there are the
\emph{init} and \emph{load\_once} functions
that are called once at the beginning of the loop to initialize local variables and
possibly perform calculations that have to be performed only once for the whole loop. 
The difference between \emph{init} an \emph{load\_once} is that \emph{init} is called exactly once for the whole
loop while \emph{load\_once} is called for every time the loop is unrolled. The
\emph{cleanup} function complements \emph{init}. Some expression nodes contain an additional \emph{reduction} function
which is used to reduce multiple local variables to a single result.
In addition to the functions defining the operation to be performed the loop may require local
variables for intermediate results. When the loop is unrolled it might even
require one set of variables for each time it was unrolled. Additionally there can
be variables that are only needed once for the whole loop despite the unrolling.
Since the loop template cannot declare an arbitrary number of local variables it
instead declares a fixed amount of variables (one for every time the loop is unrolled)
with a composite type \emph{Storage} that is also provided by the expression tree.
Similarly there is a type \emph{TemporaryStorage} that is only instantiated once
by the loop template. Listing \ref{protonode} shows the basic interface of a
expression node and listing \ref{looptemplate} shows an example loop template that
is unrolled twice.

\begin{program}
{\footnotesize 
\begin{verbatim}
class VExprOperation : public VExpr< VExprOperation< A,B > >
{
public:
	typedef typename A::Type Type;
	
    typedef pair<typename A::Storage, typename B::Storage>
			Storage;
			
    typedef pair<typename A::TemporaryStorage, typename B::TemporaryStorage>
			TemporaryStorage;

    void init(TemporaryStorage &ts) const;
    
    void cleanup(TemporaryStorage &ts) const;
    
    void load_once(Storage &s, TemporaryStorage &ts) const;
    
    void load(const int &i, Storage &s, TemporaryStorage &ts) const;
    
    void store(const int &i, Storage &s, TemporaryStorage &ts) const;
    
    Type single_op(const int &i, TemporaryStorage &ts) const
    
    typename vectorizer<Type>::vector_t
    vector_op(const int &i, Storage &s, TemporaryStorage &ts) const;
};
\end{verbatim}
}
\caption{The interface of a expression template node.}
\label{protonode}
\end{program}

\begin{program}
{\footnotesize 
\begin{verbatim}
    static void run(const VExpr<A>& a)
    {
        const A& ao ( a );

        typedef vectorizer<typename A::Type> vec;
        int i = 0;
        int len = ao.size();

        typename A::TemporaryStorage ts;

        ao.init(ts);

        if(vec::specialized)
        {
            const int n = len&(~(2*vec::len-1));

            typename A::Storage s1;
            typename A::Storage s2;

            ao.load_once(s1, ts);
            ao.load_once(s2, ts);

            for(; i<n; i+=2*vec::len)
            {
                ao.load(i,            s1, ts);
                ao.load(i+  vec::len, s2, ts);

                ao.vector_op(i,            s1, ts);
                ao.vector_op(i+  vec::len, s2, ts);

                ao.store(i,            s1, ts);
                ao.store(i+  vec::len, s2, ts);
            }
        }

        for(; i<len; ++i)
        {
            ao.single_op(i, ts);
        }
        ao.cleanup(ts);
    }
\end{verbatim}
}
\caption{Example loop template that shows a two times unrolled loop}
\label{looptemplate}

\end{program}

\subsection{Loop Structure}
To reach near optimal performance the loop has to be structured so it minimizes
pipeline stalling. This can be achieved by using as many registers as possible
and by interleaving instructions such that the distances between the usage of
each individual register is maximized. This is done by grouping instructions into packages
and starting each instruction package with a burst of load commands followed by a burst of operation commands and
finishing with a burst of store commands, while retaining the relative order of
variable/register usage inside each burst. This means that by the time the
operation command is called on the first register the load command has had the maximum
possible amount of time to complete. Also grouping load and store instructions
helps the CPU to optimize memory bus usage.

To ensure optimal register usage the instruction package size has to be chosen depending
on the amount of registers used by each operation. This number can be retrieved
from the expression tree and the appropriate loop template can be chosen at compile
time. For that purpose the \emph{run} function as seen in listing \ref{looptemplate}
is a member function of a template struct that is specialized for different
register usage patterns.

A single iteration of the unrolled loop can contain multiple instruction packages.
The number of packages per loop was determined by trial. Choosing low (usually
only one) amounts of packages gives high performance for small vector sizes while
higher amounts yield better performance for large vectors. We chose high amounts for SALT
since those approximate the performance of BLAS implementations better.

\section{Performance Benchmarks}
\begin{figure}
\includegraphics[ width=\textwidth ]{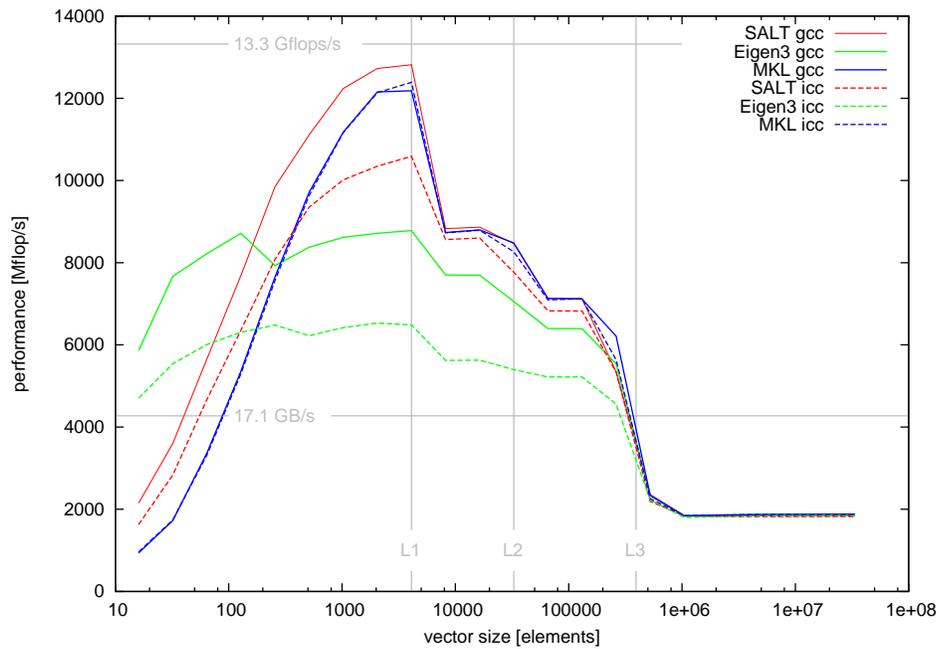}
\caption{Performance results for the single precision dot product (sDOT) on a Intel Core i5-580M CPU.
The grey lines indicate the sizes of the different caches and the 17.1 GB/s memory bandwidth and
13.3 Gflop/s theoretical peak performance of the used processor.}

\label{i5sdot}
\end{figure}

\begin{figure}
\includegraphics[ width=\textwidth ]{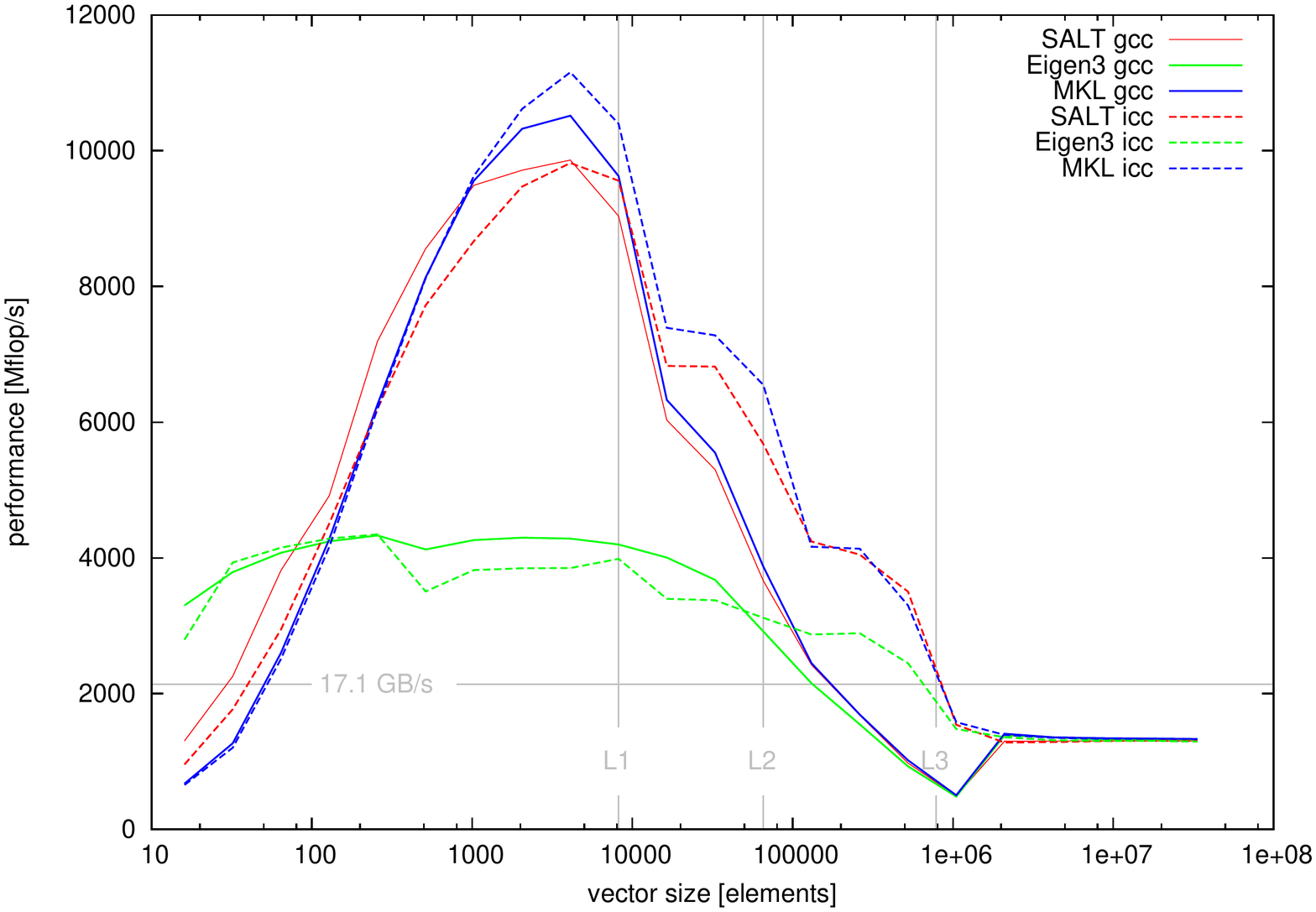}
\caption{Performance results for the single precision vector scaling operation (sSCAL) on a Intel Core i5-580M CPU.
The grey lines indicate the sizes of the different caches and the 17.1 GB/s memory bandwidth and
13.3 Gflop/s theoretical peak performance of the used processor.}
\label{i5sscal}
\end{figure}

\begin{figure}
\includegraphics[ width=\textwidth ]{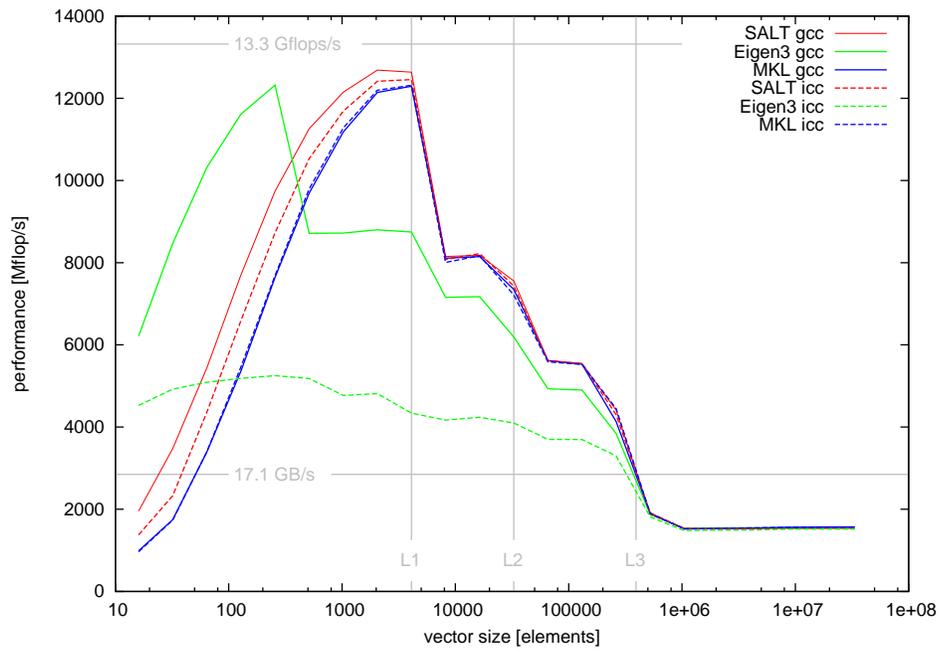}
\caption{Performance results for the single precision operation $y = y + \alpha x$ (sAXPY) on a Intel Core i5-580M CPU.
The grey lines indicate the sizes of the different caches and the 17.1 GB/s memory bandwidth and
13.3 Gflop/s theoretical peak performance of the used processor.}
\label{i5saxpy}
\end{figure}

\begin{figure}
\includegraphics[ width=\textwidth ]{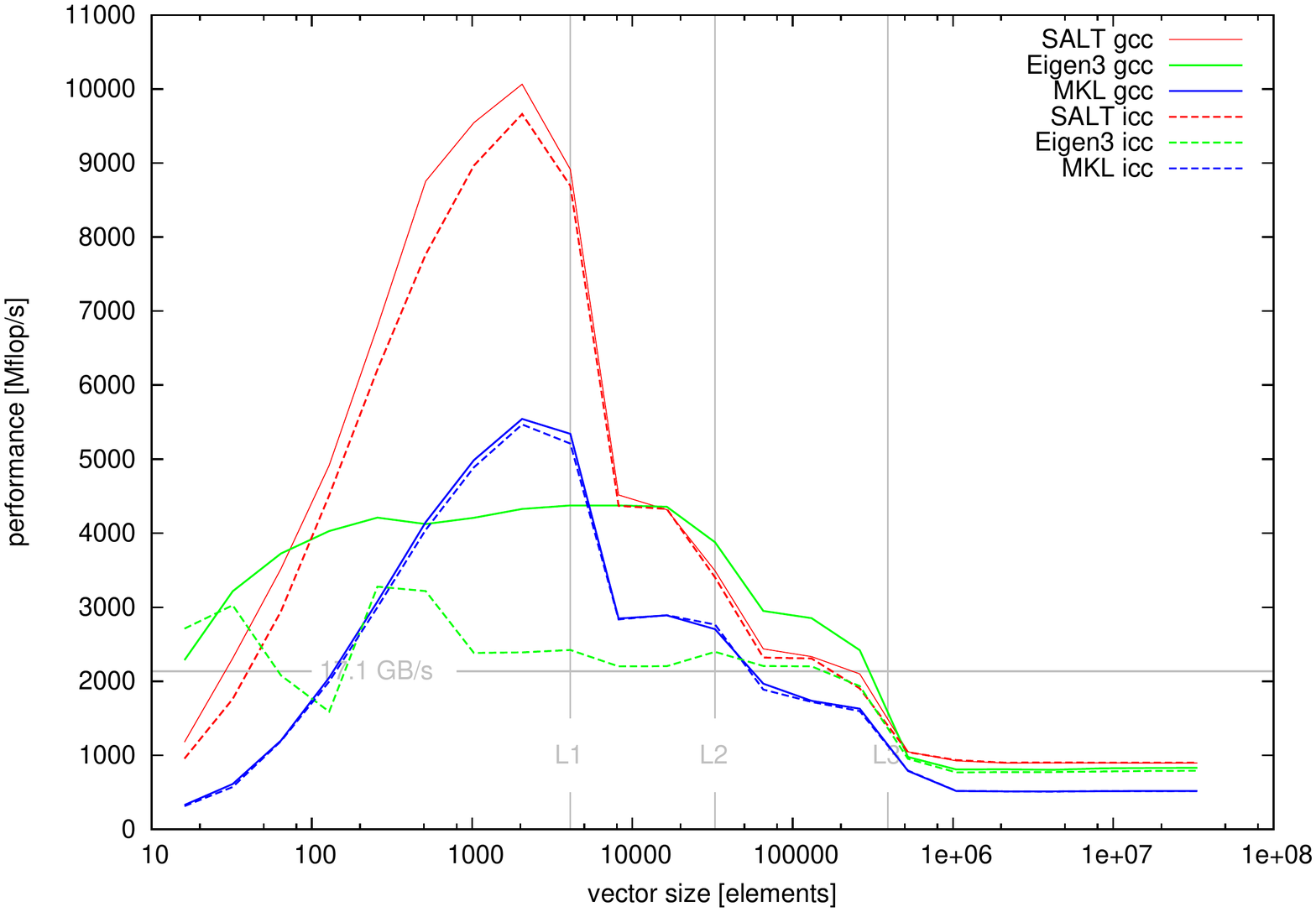}
\caption{Performance results for the single precision out-of-place vector scaling operation (sCOPY + sSCAL) on a Intel Core i5-580M CPU.
The grey lines indicate the sizes of the different caches and the 17.1 GB/s memory bandwidth and
13.3 Gflop/s theoretical peak performance of the used processor.}
\label{i5sscal2}
\end{figure}

Measurements were taken for three common vector operations. These are the dot product
(xDOT), vector scaling (xSCAL) and operations of the form $y = y + \alpha x$
(xAXPY). The results were compared to Intel's MKL library (BLAS interface) and
to Eigen (expression templates). The results for a Intel Core i5-580M CPU (2.66 GHz,
3.33 GHz TurboBoost) are shown in figures \ref{i5sdot}, \ref{i5sscal} and \ref{i5saxpy}.
Additionally figure \ref{i5sscal2} shows the results for a out-of-place vector
scaling operation which cannot be expressed as a single BLAS call and therefore
has to be written as combination of a copy and a scaling operation when using BLAS
but gets compiled into a single loop by the expression templates.
Compiler version and optimization flags for both compilers are shown in table \ref{cflags}.

\begin{table}[h]
\begin{center}
\begin{tabular}{| l | l | l | }
  \hline                       
  Compiler & Version & Flags \\
  \hline                       
  GCC & 4.5.2 & \texttt{-O3 -msse3} \\
  ICC & 12.0.2 & \texttt{-O3 -msse3} \\
  \hline  
\end{tabular}
\end{center}
\caption{Compiler versions and flags.}
\label{cflags}
\end{table}


\section{Higher Level Parallelisation}
The usage of SIMD operations already provides instruction level parallelism, but
one might be interested in additional parallelization on the thread and process level.
Thread level parallelization which is attractive for modern multicore architectures
could be easily achieved by inserting OpenMP pragmas into the loop templates. In
that case care has to be taken to avoid false sharing and excessive overhead from
spawning threads. For small sized vectors that fit into the L1 cache of the processor
the execution times of the whole operation is in the range a few hundred thousand cycles.
Spawning a thread each time the operation is executed will often have the 
opposite effect of reducing performance by introducing overhead. In case of very
large vectors it has to be noted that the rate at which even a single core is
able to process data usually exceeds the memory bandwidth of the system. The
sDOT (figure \ref{i5sdot}) example demonstrates this by achieving a peak throughput of about 60 GB/s
for vectors that completely fit into the L1 cache while the maximum memory bandwidth
of the used processor amounts to only 17.1 GB/s. For cases where the operation
is already limited by memory bandwidth on a single core, using multiple cores
that share a memory bus will not increase performance. Better results might
be achieved by parallelizing outside of the vector library.

Process level parallelization using a data parallel ansatz and MPI communication,
which is popular in software for cluster computers, can also be provided by a 
expression template library. But since the data parallel ansatz doesn't directly
affect how the individual operations are carried out in each process, it is best
dealt with by an additional abstraction layer thus enforcing the single
responsibility principle.

\section{Conclusions}
The benchmarks show that our new approach - Statically Accelerated Loop Templates (SALT) - allows template libraries to match the
performance of BLAS libraries and even outperform them in cases that require
the composition of BLAS calls. Performance inconsistencies across different
compilers are greatly reduced in comparison to to existing template libraries.
It retains the math-like syntax and better integration into standard C++ that
comes with using C++ specific features like operator overloading and generic
programming, and allows existing template algorithms to instantly benefit
from efficient vectorization.
The strong separation of low level instructions, expression building and
instruction ordering into vectorizer class, expression nodes and loop templates
gives unique access points for each aspect of the algorithms and therefore
simplifies customization and extension of the framework by following the single
responsibility principle.
Additionally the pure template character of the library makes it easy to use
and lightweight since no additional libraries have to be linked or compiled.

\bibliography{paper.bib}{}
\bibliographystyle{unsrt}

\end{document}